\documentclass[aps,prd,preprint,superscriptaddress]{revtex4}

\usepackage{latexsym}
\usepackage{amsmath,amssymb}

\begin{document}


\begin{flushright}
{\tt KIAS-P12059}
\end{flushright}

\title{Nonextremal Kerr/CFT on a stretched horizon}


\author{Ee Chang-Young} %
\email[]{cylee@sejong.ac.kr} %
\affiliation{Department of Physics and Institute of Fundamental
  Physics, Sejong University, Seoul, 143-747, Korea} %

\author{Myungseok Eune} %
\email[]{younms@sogang.ac.kr} %
\affiliation{Research Institute for Basic Science, Sogang University,
  Seoul, 121-742, Korea} %

\date{\today}

\vspace{20mm}

\begin{abstract}
  We study the conformal symmetry described by $SL(2,R)_L \times
  SL(2,R)_R$ in the nonextremal Kerr black hole case. We calculate the
  central charges $c_{L,R}$ separately on a stretched horizon by the
  Hamiltonian gravity.  In order to get two sets of conformal
  symmetry, we consider two Killing vectors each of which depends on either
  one of the two independent coordinates separated in
  the spirit of point-splitting from the angular
  coordinate of the corotating frame at the horizon.   Then, we obtain the
  Bekenstein-Hawking entropy by the Cardy formula.
\end{abstract}


\maketitle

\section{Introduction}
\label{sec:intro}

Since the successful description of the holographic duality between
the geometry near the horizon of extremal Kerr black hole and the
conformal field theory~\cite{Guica:2008mu, Matsuo:2009pg,
  Matsuo:2009sj, Bredberg:2011hp, Chen:2011wt}, there have been many
studies for the holographic duality in various black
holes~\cite{Compere:2008cv, Compere:2009zj, Blagojevic:2009ek,
  Hartman:2008pb, Compere:2009dp, Chen:2009ht, Peng:2009wx,
  Garousi:2009zx, Li:2010ch, Matsuo:2010ut, Guica:2010ej,
  Rasmussen:2010sa, Henneaux:2011hv, Goldstein:2011jh,
  Azeyanagi:2011zj, Guneratne:2012qp, Button:2010kg,
  Rodriguez:2010ev}.  The holographic duality has been also developed
for near extremal black holes by introducing the extremal
parameter~\cite{Matsuo:2010ut, Rasmussen:2010sa} or calculating the
graybody factor~\cite{Bredberg:2009pv, Cvetic:2009jn, Hartman:2009nz,
  Chen:2010bsa}. And the hidden conformal field theory described by
the $SL(2,R)_L \times SL(2,R)_R$ in the Kerr black hole has been
studied from the scalar wave equation~\cite{Castro:2010fd}.
However, in this case the values of the central charges $c_{L,R}$ have
been just assumed to be the same as in the extremal black hole case.
Up to now, the left and right central charges $c_{L,R}$ have been
calculated only in the near extremal case~\cite{Matsuo:2010ut}.
There have been many extended studies of the hidden conformal symmetry
for various black holes~\cite{Fareghbal:2010yd, Setare:2011zzg,
  Chen:2011gz, Chen:2011kt, Franzin:2011wi, Lowe:2011aa,
  Bertini:2011ga, Setare:2011zze, Setare:2011zzi, Chen:2012np,
  Ghezelbash:2012qn, Cvetic:2011hp, Dias:2012pp}, but no calculation
of the both central charges has been done. Recently,
 for the nonextremal Kerr black hole, the central charge has been obtained
successfully by using the Hamiltonian gravity on a stretched horizon,
but it describes only $SL(2,R)_L$ symmetry, not $SL(2,R)_L \times
SL(2,R)_R$~\cite{Carlip:1998wz, Park:1999tj, Park:2001zn,
  Carlip:2011ax, Carlip:2011vr}.
There also have been studies for extremal black holes using the
Hamiltonian gravity~\cite{Chen:2011wm, Mei:2012wd}.

In this paper, we investigate the conformal symmetry described by $SL(2,R)_L
\times SL(2,R)_R$ on a stretched horizon in the nonextremal Kerr black
hole case. Adopting the method of Carlip~\cite{Carlip:2011ax,
  Carlip:2011vr},
we obtain the central charges $c_{L,R}$ separately.
The paper is organized as follows. In
Sec.~\ref{sec:rev.HG}, we review briefly the Hamiltonian gravity. In
Sec.~\ref{sec:nonextr-kerr-black}, we obtain the central charges
$c_{L,R}$, the temperatures $T_{L,R}$, and the total entropy $S$ on a
stretched horizon in the nonextremal case with the Hamiltonian
gravity.
In order to have two Virasoro algebras we consider two Killing vectors
which are functions of either one of the two coordinates split from
the angular coordinate of the corotating frame at the horizon.  This
splitting was done in the spirit of point-splitting.  Then the central
charges and the total entropy are evaluated.
Finally, in Sec.~\ref{sec:discus}, we conclude with discussion.

\section{Brief review on Hamiltonian gravity}
\label{sec:rev.HG}

In order to obtain the central charges, we evaluate the central terms
in the Hamiltonian gravity in this work.  Here, we briefly review the
related ingredients in the Hamiltonian gravity for this
calculation. The detailed review can be found in
Ref.~\cite{Carlip:2011ax}.  We consider a stationary nonextremal black
hole, whose metric can be written in the ADM form as follows.
\begin{align}
  ds^2 &= -N^2 dt^2 + q_{ij} (dx^i + N^i dt)(dx^j + N^j
  dt). \label{metric:ADM}
\end{align}
Under a diffeomorphism generated by a vector field $\xi^\mu$, the
metric~\eqref{metric:ADM} transforms as
\begin{align}
  \delta_{\xi} N &= \bar{\partial}_t \xi^{\perp} + \hat{\xi}^i \partial_i
  N, \label{diff:N} \\
  \delta_{\xi} N^i &= \bar{\partial}_t \hat{\xi}^i - N
  q^{ij} \partial_j \xi^{\perp} + q^{ik} \partial_k N \xi^{\perp} +
  \hat{\xi}^j \partial_j N^i, \label{diff:N:i}  \\
  \delta_{\xi} q_{ij} &= q_{ik} \left( \partial_j \hat{\xi}^k -
    \frac{\partial_j N^k}{N} \xi^{\perp} \right) + q_{jk}
  \left(\partial_i \hat{\xi}^k - \frac{\partial_i N^k}{N} \xi^\perp
  \right) + \frac{1}{N} \xi^\perp \bar{\partial}_t q_{ij} +
  \hat{\xi}^k \partial_k q{ij}, \label{diff:q:ij}
\end{align}
where the convective derivative $\bar\partial_t$ is defined by
\begin{align}
  \bar{\partial}_t &\equiv \partial_t -
  N^i \partial_i, \label{def::bar:dt}
\end{align}
and the Killing vector $\xi^\mu$ is redefined as
\begin{align}
  \xi^\perp &\equiv N \xi^t, \qquad \hat{\xi}^i \equiv \xi^i + N^i
  \xi^t, \label{killing:SD}
\end{align}
which are called the ``surface deformation parameters''.  The
symmetries of canonical general relativity are generated by the
Hamiltonian
\begin{align}
  H[\xi] = \int d^3 x \left(\xi^\perp {\cal H} + \hat\xi^i {\cal H}_i
  \right), \label{H}
\end{align}
with
\begin{align}
  {\cal H} = \frac{1}{\sqrt{q}} (\pi^{ij} \pi_{ij} - \pi^2) - \sqrt{q}
  {\cal R}, \qquad {\cal H}^i = -2 D_j \pi^{ij}, \label{H:P::constraints}
\end{align}
where $q_{ij}$, $\pi^{ij}$, and $\cal R$ are the spatial metric, the
canonical momentum, and the spatial curvature scalar, respectively,
and $D_i$ is the spatial covariant derivative with respect to
$q_{ij}$. $\cal H$ and ${\cal H}^i$ are the Hamiltonian and momentum
constraints, respectively.  In Refs.~\cite{Brown:2000dz,
  Carlip:2011ax}, a new generator with a well-defined variation and
with no boundary terms is defined by
\begin{align}
  \bar{H}[\xi] = H[\xi] + B[\xi], \label{H:bar}
\end{align}
where $B[\xi]$ depends only on the fields and parameters at the
boundary, and is chosen to cancel the boundary terms in the variation
of $H[\xi]$. The Poisson bracket of $\bar{H}[\xi]$ can be written as
\begin{align}
  \{ \bar{H}[\xi], \bar{H}[\eta] \} = \bar{H}[\{\xi, \eta\}_{\rm SD}] +
  K[\xi,\eta], \label{PB:H,H}
\end{align}
where the central term is given by
\begin{align}
  K[\xi, \eta] = B[\{\xi, \eta\}_{\rm SD}] &- \frac{1}{8\pi G}
  \int_{\partial \Sigma} d^2 x \sqrt{\sigma} n^k \bigg[
  \frac{1}{\sqrt{q}} \pi_{ik} \{\xi, \eta\}_{\rm SD}^i -
  \frac{1}{2\sqrt{q}} (\hat\xi_k \eta^\perp - \hat\eta_k \xi^\perp)
  {\cal H} \notag \\
  &+ (D_i \hat\xi_k D^i \eta^\perp - D_i \hat\eta_k D^i \xi^\perp) -
  (D_i \hat\xi^i D_k \eta^\perp - D_i \hat\eta^i D_k \xi^\perp) \notag
  \\
  &+ \frac{1}{\sqrt{q}} \left(\hat\eta_k \pi^{mn} D_m \hat\xi_n -
    \hat\xi_k \pi^{mn}D_m \hat\eta_n \right) \bigg]. \label{K:def}
\end{align}
Here, the surface deformation brakets are given by
\begin{align}
  \{\xi, \eta\}_{\rm SD}^\perp &= \hat\xi^i D_i \eta^\perp -
  \hat\eta^i  D_i \xi^\perp, \notag \\
  \{\xi, \eta\}_{\rm SD}^i &= \hat\xi^k D_k \hat\eta^i - \hat\eta^k
  D_k \hat\xi^i + q^{ik} \left(\xi^\perp D_k \eta^\perp - \eta^\perp
    D_k \xi^\perp \right).
  \label{SD:def}
\end{align}
We can calculate the central charges from the central term $K[\xi,
\eta]$ given by Eq.~\eqref{K:def}.  When a Virasoro subalgebra of the
group of surface deformation is found in the following form, $\{\xi,
\eta\}_{\rm SD} = \xi \eta' - \eta \xi'$ where the prime denotes the
derivative with respect to the rotating angle $\varphi$, then the
boundary contribution $B[\{\xi, \eta\}_{\rm SD}]$ can be ignored for
our purpose~\cite{Carlip:2011ax}.  This is because that the central
charge of the Virasoro algebra is given by the coefficient of the
 expression $\int d\varphi (\xi' \eta'' - \eta' \xi'')$
 extracted from $K$ in Eq.~\eqref{K:def}.

\section{CFT in nonextremal rotating black hole}
\label{sec:nonextr-kerr-black}

We now consider the nonextremal Kerr black hole described by
\begin{align}
  ds^2 = - \frac{\Delta}{\Sigma} \left( dt - a \sin^2 \theta\,
    d\varphi \right)^2 + \frac{\Sigma}{\Delta} dr^2 + \Sigma \,
  d\theta^2 + \frac{\sin^2\theta}{\Sigma} \left[a dt - (r^2 + a^2)
    d\varphi \right]^2, \label{metric:BL}
\end{align}
in the Boyer-Lindquist coordinates, with
\begin{align}
  \Delta(r) &= r^2 + a^2 - 2M r, \label{def:Delta} \\
  \Sigma(r,\theta) &= r^2 + a^2 \cos^2\theta, \label{def:Sigma}
\end{align}
where $M$ is the mass of the black hole and $a$ is a parameter related
to the angular momentum $J$ of the black hole as $J = a M$. The
nonextremal Kerr black hole \eqref{metric:BL} can be written in the
ADM form as
\begin{align}
  ds^2 &= - N^2 dt^2 + d\rho^2 + q_{\varphi\varphi} (d\varphi +
  N^\varphi dt)^2 + q_{\theta\theta} d\theta^2, \label{metric:ADM:K}
\end{align}
where $\rho$ is the proper distance from the horizon. In general, the
line element of the nonextremal stationary rotating black hole can be
expressed as in Eq.~\eqref{metric:ADM:K}. From now on, we
consider the nonextremal stationary rotating black hole in four
dimensions. Near the horizon of the black hole, the lapse function $N$
and the shift vector $N^\varphi$ in the metric~\eqref{metric:ADM:K}
can be expanded as
\begin{align}
  N &= \kappa_{\rm H} \rho + \frac{1}{3!} \kappa_2(\theta) \rho^3 +
  \cdots, \label{N:expand} \\
  N^\varphi &= -\Omega_H - \frac12 \omega_2(\theta) \rho^2 +
  \cdots, \label{N:phi:expand}\\
  q_{\varphi\varphi} &= q_{\varphi\varphi}^{(H)} (\theta) + \frac12
  q^{(2)}_{\varphi\varphi} (\theta) \rho^2 +
  \cdots, \label{q:phiphi:expand} \\
  q_{\theta\theta} &= q_{\theta\theta}^{(H)} (\theta) + \frac12
  q^{(2)}_{\theta\theta} (\theta) \rho^2 +
  \cdots, \label{q:thetatheta:expand}
\end{align}
where $\kappa_{\rm H}$ and $\Omega_H$ are the surface gravity and the
angular velocity on the event horizon, respectively. For convenience,
we introduce a parameter $\varepsilon$ defined by $\varepsilon
\equiv - \frac12 \omega_2 (\theta) \rho^2$ so that $N^\varphi
\approx -\Omega_H + \varepsilon$ up to the order of $\rho^2$. The shift
vector $N^\varphi$ becomes the angular velocity near horizon, when
$\rho \to 0$.

When the black hole geometry is given by Eq.~(\ref{metric:ADM:K}), the
diffeomorphism given by (\ref{diff:N})-(\ref{diff:q:ij}) becomes
\begin{align}
  \xi^\rho &= -\rho \bar\partial_t \xi^t, \label{diff:1} \\
  \bar\partial_t \xi^\rho &= \kappa_{\rm H}^2 \rho^2 \partial_\rho
  \xi^t, \label{diff:2} \\
  \bar\partial_t \hat\xi^\varphi &= \kappa_{\rm H}^2 \rho^2
  q^{\varphi\varphi} \partial_\varphi \xi^t -
  \frac{2\varepsilon}{\rho} \xi^\rho, \label{diff:3} \\
 \partial_\rho \hat\xi^\varphi &= -
 q_{\varphi\varphi}^{-1} \partial_\varphi \xi^\rho +
 \frac{2\varepsilon}{\rho} \xi^t,  \label{diff:4}
\end{align}
where the convective derivative $\bar\partial_t$ in
Eq.~\eqref{def::bar:dt} becomes
\begin{align}
  \bar{\partial}_t = \partial_t - N^\varphi \partial_\varphi
  \approx \partial_t + (\Omega_H -
  \varepsilon) \partial_\varphi. \label{bar:dt:phi}
\end{align}

Note that \eqref{metric:ADM:K} can be rewritten as
\begin{align}
  ds^2 &= q_{\varphi\varphi} (d\varphi - \Omega_+ dt) (d\varphi -
  \Omega_- dt) + d\rho^2 + q_{\theta\theta} d\theta^2,
\end{align}
where
\begin{align}
  \Omega_\pm \equiv -N^\varphi \pm
  \frac{N}{\sqrt{q_{\varphi\varphi}}}. \label{def:Omega:max:min}
\end{align}
Introducing a new parameter,
\begin{equation}
\bar\varepsilon \equiv \kappa_{\rm H} \rho /
\sqrt{q_{\varphi\varphi}},
\label{baresp}
\end{equation}
we can write $\Omega_\pm$  as
\begin{equation}
  \Omega_\pm = \Omega_H - \varepsilon
  \pm \bar\varepsilon, \label{angvel}
\end{equation}
up to the order of $\rho^2$.

 We choose our stretched horizon such that
its maximum and minimum angular velocities are given by
$\Omega_+$ and $\Omega_-$, respectively, defined by \eqref{angvel}.
One can easily check that our new Killing vectors $\chi_\pm
\equiv \partial_t + \Omega_\pm \partial_\varphi$ are null.
 For the purpose of studying the conformal field theory, we introduce a
new coordinate system defined by
\begin{align}
  \varphi_\pm = 2(\varphi - \Omega_\pm t). \label{def:new:coord}
\end{align}
We assign the coordinates $\varphi_+$ and $\varphi_-$ as the right
and left conformal coordinates, respectively.
The two coordinates are slightly separated from the angular coordinate
of the corotating frame at the horizon, $ \bar{\varphi} = \varphi -
\Omega_H t$.
We may consider these two coordinates as two-fold split of the angular
coordinate of the corotating frame at the horizon in the spirit of
point-splitting \cite{schnabl2000}.

Now, we assume the existence of two Killing vectors, which are
functions of either $\varphi_+$ or $\varphi_-$, dubbed as the right
($+$) and left ($-$) handed Killing vectors.
We thus consider the right/left handed conformal field theory (R/LCFT)
where the Killing vectors $\xi_\pm$ are the functions of $\varphi_\pm$,
respectively.  In that case, the expressions for the diffeomorphism
invariance given by Eqs.~\eqref{diff:1}, \eqref{diff:2},
\eqref{diff:3}, and \eqref{diff:4} become
\begin{align}
  \xi_\pm^\rho &= \pm \rho \bar\varepsilon \, \partial_\varphi
  \xi_\pm^t, \\
  \rho \partial_\rho \xi_\pm^t &= -
  \frac{\bar\varepsilon^2}{\kappa_{\rm H}^2} \partial_\varphi^2 \xi_\pm^t, \\
  \hat\xi_\pm^\varphi &= \mp \bar\varepsilon\, \xi_\pm^t, \\
  \partial_\rho \hat\xi_\pm^\varphi &= \mp \rho \bar\varepsilon\,
  q^{\varphi\varphi} \partial_\varphi^2 \xi_\pm^t +
  \frac{2\varepsilon}{\rho} \xi_\pm^t,
\end{align}
up to the leading order.  From Eq.~\eqref{K:def}, the terms that
contribute to the central charge are given by
\begin{align}
  K[\xi_\pm, \eta_\pm] &= \pm \frac{\bar\varepsilon^3}{8\pi G
    \kappa_{\rm H} } \cdot \frac{A}{2\pi} \int d\varphi
  \left(\partial_\varphi \xi_\pm^t \partial_\varphi^2 \eta_\pm^t
    - \partial_\varphi \eta_\pm^t \partial_\varphi^2 \xi_\pm^t
  \right),
  \label{R:K:final}
\end{align}
where $A$ denotes the area of the stretched horizon.

We obtain the surface deformation brackets~\eqref{SD:def} for the generators
up to the leading order as follows.
\begin{align}
  \{\xi_\pm, \eta_\pm \}_{\rm SD}^t &= \mp 2 \bar\varepsilon
  (\xi_\pm^t \partial_\varphi \eta_\pm^t - \eta_\pm^t \partial_\varphi
  \xi_\pm^t), \notag \\
  \{\xi_\pm, \eta_\pm \}_{\rm SD}^\rho &= \pm \rho
  \varepsilon\, \partial_\varphi \{\xi_\pm, \eta_\pm \}_{\rm SD}^t ,
  \notag \\
  \{\xi_\pm, \eta_\pm \}_{\rm SD}^\varphi &= \mp \varepsilon
  \{\xi_\pm, \eta_\pm \}_{\rm
    SD}^t. \label{xi:eta:bracket4normalization}
\end{align}
 The commutator between the right and left handed
generators are obtained as
\begin{align}
  \{\xi_+, \eta_- \}_{\rm SD}^t = \{\xi_+, \eta_- \}_{\rm SD}^\rho =
  \{\xi_+, \eta_- \}_{\rm SD}^\varphi =
  0, \label{xi:eta:orthogonality}
\end{align}
up to the leading order. Here we would like to  note
 that \eqref{xi:eta:orthogonality} is obtained when  our new parameter $\bar\varepsilon$
 is defined by \eqref{baresp}.
Eq.~\eqref{xi:eta:orthogonality} shows that the two Virasoro algebras of $\xi_+$
and $\xi_-$ are independent of each other, and the two Virasoro
algebras contribute separately to the entropy.  One can also see from
\eqref{xi:eta:bracket4normalization} that
$\xi_\pm^t$ have to be  normalized as
\begin{align}
  \xi_\pm^t = \mp \frac{1}{2\bar\varepsilon} \,
  \tilde\xi_\pm^t, \label{xi:normalized}
\end{align}
where the normalized ones are denoted by $\tilde\xi_\pm^t$.  Taking
the above normalization into account, the central charge can be read
as
\begin{align}
  c_{R/L} \equiv c_\pm = \frac{3\bar\varepsilon A}{2\pi G \kappa_{\rm
      H}}. \label{R:c}
\end{align}

The temperatures, both the right and left handed ones, can be
calculated by comparing the Boltzmann factor.  For a scalar field in
the Frolov-Thorne vacuum, the wave function in the eigenmodes of
energy $\omega$ and angular momentum $m$ can be written as $\Phi \sim
e^{-i\omega t + i m \varphi}$. Then, using the relation
\begin{align}
  e^{-i\omega t + i m \varphi} = e^{-i n_- \varphi_- + i n_+
    \varphi_+}, \label{FTvacuum:scalar:wave}
\end{align}
we obtain the eigenmodes in the new coordinate system of $
\varphi_\pm$ as follows
\begin{align}
  n_\pm &= \frac{1}{4\bar\varepsilon}(\omega - m \Omega_\mp).
\end{align}
In terms of these variables, the Boltzmann factor can be written as
\begin{align}
  e^{-(\omega-m\Omega_H)/T_H} = e^{- n_+/T_+ -
    n_-/T_-}, \label{Boltzmann}
\end{align}
where the dimensionless left and right temperatures are given by
\begin{align}
  T_{R/L} \equiv T_\pm &= \frac{T_H}{2\bar\varepsilon}, \label{T:R:L}
\end{align}
with $T_H =\kappa_{\rm H}/2\pi$.

Finally, we calculate the entropy with the help of the Cardy formula.
From Eqs.~\eqref{R:c} and~\eqref{T:R:L}, the entropy of the right/left
part is given by
\begin{align}
  S_{R/L} = \frac{\pi^2}{3} c_{R/L} T_{R/L} =
  \frac{A}{8G}. \label{S:R:L}
\end{align}
Thus we obtain the total entropy of the system as
\begin{align}
  S = S_R + S_L = \frac{A}{4G}, \label{S:total}
\end{align}
which agrees with the Bekenstein-Hawking entropy.

\section{Conclusion}
\label{sec:discus}

In this paper, we obtain the entropy of the nonextremal Kerr black
hole by considering two copies of the Virasoro algebras.  In order to
have two copies of the Virasoro generators, we introduce the left and
right handed coordinates that correspond to
the corotating coordinates of the minimum and maximum angular
velocities of the stretched horizon, respectively, by adopting the
 point-splitting idea.
 We choose  our new parameter
$\bar\varepsilon$ such that the surface deformation brackets between the two Virasoro
generators commute. Showing their independence,
 we calculate the central charges of the
two Virasoro algebras and the corresponding entropies independently using the Cardy
formula.  In our setup, each part contributes the
same amount of entropy, $\frac{A}{8G}$, making the total entropy
agree with the Bekenstein-Hawking entropy.

Now we make a final comment on the null condition for
the Killing vectors on the stretched horizon.
In Refs.~\cite{Carlip:2011ax,Carlip:2011vr}, $\bar\varepsilon$ has
been chosen such that the Killing vector $\bar\chi^\mu
= \partial_t^\mu + \bar\Omega \partial_\varphi^\mu$ is null on the
stretched horizon.  Notice that $\bar\Omega$ agrees with our $\Omega_-$
up to the first order in $\rho$. There the condition was set by
imposing the null condition of the Killing vector up to the leading
order in $ \rho$:
\begin{align}
  0 &= \bar\chi^2 = -N^2 + q_{\varphi\varphi} (N^\varphi +
  \bar\Omega)^2 \notag \\
  & = -\kappa_{\rm H}^2 \rho^2 + q_{\varphi\varphi}^{(H)}
  \bar\varepsilon^2 + O(\rho^3).
\end{align}
Thereby $\bar\varepsilon$ is set to satisfy $\bar\varepsilon^2 =
\kappa_{\rm H}^2 \rho^2 / q_{\varphi\varphi}^{(H)} $,
which agrees with our definition.
However, if we replace $\bar\Omega$
 with $\Omega_\pm$ in the above defining relation of $\bar\chi$,
we end up with our new Killing vectors,
 $\chi_\pm = \partial_t +
\Omega_\pm \partial_\varphi$,
which satisfy the null condition up to the subleading order in
$\rho$.


\begin{acknowledgments}
  ECY would like to thank KIAS for hospitality during the time that
  this work was done.  This work was supported by National Research
  Foundation(NRF) of Korea Grants funded by the Korean Government
  (Ministry of Education, Science and Technology),
  NRF-2010-359-C00007(ME) and NRF-2011-0025517(ECY).
  We thank the referee for critically important comments that led to 
  essential improvement of the paper.
\end{acknowledgments}



\end{document}